\documentclass[12pt,preprint]{aastex}

\usepackage{emulateapj5}
\usepackage{epsf}
\usepackage{graphics}

\newenvironment{inlinefigure}{
\def\@captype{figure}
\noindent\begin{minipage}{0.999\linewidth}\begin{center}}
{\end{center}\end{minipage}\smallskip}

\slugcomment{Accepted to the Astrophysical Journal}

\shorttitle{GRB Galaxy Morphology}
\shortauthors{C.J. Conselice et al.}

\def\deg{$^{\circ}\,$}

\begin{document}

\title{Gamma-Ray Burst Selected High Redshift 
Galaxies: Comparison to Field Galaxy Populations to $z \sim 3$}

\author{C. J. Conselice\altaffilmark{1,2},
P. M. Vreeswijk\altaffilmark{3},
A. S. Fruchter\altaffilmark{4},
A. Levan\altaffilmark{4,5},
C. Kouveliotou\altaffilmark{6},
J.P.U. Fynbo\altaffilmark{7},
J. Gorosabel\altaffilmark{4,8},
N.R. Tanvir\altaffilmark{9},
S. E. Thorsett\altaffilmark{10}}

\altaffiltext{1}{California Institute of Technology, Mail Code 105-24, Pasadena
CA 91125; cc@astro.caltech.edu}
\altaffiltext{2}{National Science Foundation Astronomy \& Astrophysics Fellow}
\altaffiltext{3}{Astronomical Institute `Anton Pannekoek',
University of Amsterdam, \& Center for High Energy Astrophysics,
Kruislaan 403, 1098 SJ Amsterdam, The Netherlands; European Southern 
Observatory, Alonso de C\'odova 3107, Casilla 19001,  
Santiago 19, Chile}
\altaffiltext{4}{Space Telescope Science Institute, 3700 San Martin Drive,
Baltimore, MD 21218, USA; fruchter@stsci.edu}
\altaffiltext{5}{Department of Physics and Astronomy, University of Leicester, University Road, Leicester, LE1 7RH, UK}
\altaffiltext{6}{NASA/Marshall Space Flight Center, National Space Science Technology Center, 5D-50, 320 Sparkman Drive, Huntsville, AL 35805}
\altaffiltext{7}{Niels Bohr Institute, Copenhagen University, Juliane Maries Vej 30, DK-2100 Copenhagen}
\altaffiltext{8}{Instituto de Astrof\'{\i}sica de Andaluc\'{\i}a, IAA-CSIC,
   Camino Bajo de Hu\'etor, 24, E-18008 Granada, Spain}
\altaffiltext{9}{Centre for Astropysics Research,
University of Hertfordshire, College Lane, Hatfield, AL10 9AB, UK}
\altaffiltext{10}{Dept of Astronomy \& Astrophysics, University of 
California, Santa Cruz, CA 95064}

%\begin{center}
%\today
%\end{center} 
%and sizes
\begin{abstract}

  We study the internal structural properties of 37 long duration
  gamma-ray burst (GRB) host galaxies imaged with the Hubble Space Telescope.
  Our goal is to gain insights in the type of
  galaxies that gives rise to GRBs, and the relationship of GRB
  hosts to high-redshift galaxies selected through more traditional 
  photometric  methods.  We
  measure structural properties of our sample  from {\it
    Hubble Space Telescope} observations obtained after the GRB afterglow
  faded.  Fitting exponential disk (typical for spirals) and r$^{1/4}$
  (typical for ellipticals) models to the surface brightness profiles
  of eight $z < 1.2$ bright host galaxies, we find that the disk model is
  slightly preferred for most hosts, although two galaxies are fit
  best with an r$^{1/4}$ profile. We furthermore measure the central
  concentrations and asymmetries of all 37 host galaxies using the CAS (concentration, asymmetry, clumpiness)
  system, and compare with values for galaxies in the Hubble
  Deep Field, and systems present on the gamma-ray burst host
  images. Our first main conclusion is that GRB hosts exhibit a surprisingly 
  high diversity of galaxy types.  A significant fraction (68\%) of host 
  galaxies are situated in a region of the concentration-asymmetry diagram 
  occupied by spirals or peculiar/merging galaxies.  Twelve hosts
  (32\%) are situated in the region occupied by elliptical
  galaxies, having high concentration indices indicative of
  early-types or early types in formation.  These results show that GRB host galaxies are not a single
  morphological type, but span the available range of
  galaxy types seen at high redshift.    We also find some 
  evidence
  for evolution in GRB host galaxy morphology, such that hosts at
  $z > 1$ have a relatively high light concentration, indicating that these 
  systems are perhaps progenitors of massive galaxies, or are compact blue 
  star forming galaxies. We find that GRB hosts at $z > 1$ are 
  different from the
  general field population at $z > 1$ in terms of light concentration at 
  $>$99.5\% confidence, yet have sizes similar to the general $z > 1$
  galaxy population. 
  This is the opposite of the effect at $z < 1$ where GRB hosts are smaller
  than average.
  We argue that GRB hosts
  trace the starburst population at high
  redshift, as similarly concentrated galaxies at $z > 1$
  are undergoing a disproportionate amount of star formation for their
  luminosities. Furthermore,
our results show that GRBs are not only an effective tracer of star formation,
but are perhaps ideal tracers of typical galaxies undergoing star formation at
any epoch, making them perhaps our best hope of locating the earliest galaxies 
at $z > 7$.

\end{abstract}


\section{Introduction}

Galaxies at high redshift are selected in a number of different ways.
Traditional techniques include using deep optical and near infrared
imaging to select systems based on spectral features such as optical
or near infrared breaks or line emission (e.g., Steidel \& Hamilton
1992; Franx et al. 2003; Fynbo et al. 2003; Moustakas et al. 2004),
X-ray emission (Lehmer et al. 2005), sub-mm emission (e.g., Hughes et 
al. 1998) or through spectroscopic and photometric
redshift surveys, typically associated with deep {\it Hubble Space Telescope}
(HST)
imaging, or deep near-infrared imaging (e.g., Dickinson et al. 2000; 
Budav{\' a}ri et al. 2000; Stanford et al. 2004;
Somerville et al. 2004; Daddi et al. 2004).  These
techniques are all strongly biased, and are possibly
missing substantial populations of
galaxies at high redshifts.  The discovery of gamma ray bursts (GRBs)
and their associated host galaxies opens up a new possibility for
detecting and understanding the formation and evolution of galaxies,
as this selection is not based on assumptions about underlying spectral
energy distributions. It is therefore
important to understand if GRBs select different galaxy populations than
photometric methods by comparing the known properties of field galaxies 
selected through other techniques with the host galaxies of GRBs.

In practically all cases where HST observations were performed of
a GRB field, a host galaxy has been detected at the position of
the early afterglow.   These GRB host galaxies are
actively star forming, as has been demonstrated in several
different ways. First, GRB host galaxies tend to
be bluer than field galaxies at similar redshifts (Fruchter
et al. 1999, 2005; Le Floc'h et al. 2003). GRB hosts also have strong Balmer
and nebular emission lines (e.g., Kulkarni et al. 1998; Djorgovski et al.
1998) which are
produced by ionizing stars (e.g., Rhoads \& Fruchter 2001).
Moreover, there is also mounting evidence that long-duration GRBs are
associated with regions of massive-star formation within their hosts.
This has been shown through the location of afterglows
with respect to their host galaxies (e.g., Bloom et al. 2002; Fruchter
et al. 2005).  For example, Fruchter et al. (2005) argue
that GRBs are more centrally concentrated on the peaks of surface
brightness enhancements in galaxies than the total light, suggesting
a strong correlation between the locations of GRBs and bright regions
of star-formation.  Another line of evidence for a star-formation connection
was initiated by the discovery of an unusual supernova of type Ic,
SN\,1998bw, in the error box of GRB\,980425
(Galama et al. 1998), which suggested an association between
the two phenomena.  Since this discovery further observations have
found a supernova light-curve component superimposed on the
late-time power-law decay of afterglows (Bloom et al. 1999; Reichart 1999;
Galama et al. 2000; Castro-Tirado et al. 2001).  Furthermore, the system
GRB\,030329 was close enough to show spectroscopically (Stanek et al.
2003; Hjorth et al. 2003a) that at least some GRBs also produce a
special type of supernova explosion, or vice versa. This association
fits very well in the collapsar model, in which a rapidly rotating
massive star undergoing core-collapse produces a jetted GRB along the
rotation axis, and at the same time blows up the entire star in an
energetic supernova explosion (Woosley 1993; MacFadyen \& Woosley 1999).

If GRBs are caused by the explosion of a massive star, one might
expect most bursts to occur in massive star-bursting galaxies, i.e.
galaxies which are converting most of their neutral gas content into
stars in a very short period of time ($\sim$ 10$^8$ years).  However, there is
a diversity in the modes of star formation, and the kinds of
galaxies in which GRBs could originate. From radio and sub-mm observations it
appears that at most 20\% of GRB hosts are starburst galaxies,
with star-formation rates (SFRs) of several hundred M$_{\odot}$ per
year (Berger et al. 2001; Frail et al. 2002; Tanvir et al. 2004).
It has also been argued that most host galaxies are very 
blue, sub-luminous,  and sub-massive (e.g., Le Floc'h et al. 2003), although
these results might be biased by the steep luminosity function
of star forming galaxies. None of the
galaxies in the sample of Le Floc'h et al. (2003), who observed GRBs in the 
R and K bands, include known hosts with characteristics of luminous infrared/submillimeter sources
or extremely red starbursts. More recently, Christensen et al. (2004)
fitted galaxy templates to the SEDs of a
sample of hosts, finding that most have young starburst-like SEDs
with moderate to low extinction. The host galaxies of GRBs are therefore
potentially an inhomogeneous set of galaxies with ongoing star formation.

To shed light on some of these issues we investigate in this paper
the structural properties and sizes of nearly all
GRB host galaxies imaged with HST to date. Our goal is to provide additional
insight into the types of galaxies that give rise to GRBs, their
relationship to other galaxies at similar redshifts, and their
star-formation properties.  We used a sample of late-time HST images of GRB
hosts that were unambiguously identified through an accurate
projection of the early afterglow. This data set is described in \S 2.  
Our study of the morphology of these GRB
hosts is divided into two main parts: in \S 3 we fit
exponential and r$^{1/4}$ surface brightness profiles to bright hosts
at $z < 1.2$.
This allows us to make a rough distinction between early- and late-type
galaxies.  In \S 3.2 we use the central concentration and asymmetry
indices in order to classify GRB host galaxies as elliptical-like, 
spiral-like, or
peculiar. In \S 4 we discuss how GRB hosts are related to other
galaxies at high redshift, while \S 5 is a discussion of our results in
terms of previous findings concerning GRB hosts, and \S 6 is a summary of 
our results.  We assume the following cosmology in this paper: H$_0$ = 
65 km s$^{-1}$
Mpc$^{-1}$, $\Omega_{\Lambda}$ = 0.7 and $\Omega_{\rm m}$ = 0.3.

\section{The HST sample of GRB host galaxies}

We make use of observations in three HST programs that have imaged GRB
host galaxies over the past few years. The principal investigators of
these programs are: Fruchter (e.g., Fruchter et al. 2000, 2005; HST programs
7863, 8189, 9074, 9405), Kulkarni (e.g., Kulkarni et al. 1998; HST 
programs 8867, 9180), and
Holland (Holland et al. 2000; HST program 8640). We have selected the host
galaxies for which an accurate (0\farcs2) projection of the early
optical afterglow is possible, to avoid misidentifications of the host
system.  We utilize all of the imaging from these programs with the
exception of GRB~021202, GRB~020124, GRB~020321, and GRB~020531, which
are GRBs where the host galaxies was not identified.  We also exclude
an analysis of GRB~980425, a large nearby spiral galaxy, due to its small
redshift, and difficulty in retrieving structural parameters, and
GRB~000210 which has no late-time image.  We also do not consider
GRBs 980326, 990308 and 000301 in our analysis as the host galaxy
is either too faint for a morphological analysis with $R > 29$, or 
the host is not detected, or unambiguously identified.   We also
do not include GRB~000131 in our analysis, due to its extremely low
surface brightness, given its high redshift of $z = 4.5$.
It is also not known, except for GRB 000301 what redshift
these hosts are at, thus including them in our analysis and interpretations
is difficult.  It is however possible that not including these GRBs
could bias our results, although there are only four, and thus not likely
to be a significant bias. In any case, our sample is essentially
complete for bursts at $R < 27$, and we analyze our incompleteness in
\S 4.4.

\begin{figure*}
\begin{center}
\vspace{0cm}
\hspace{-1cm}
\rotatebox{0}{
\includegraphics[width=1\linewidth]{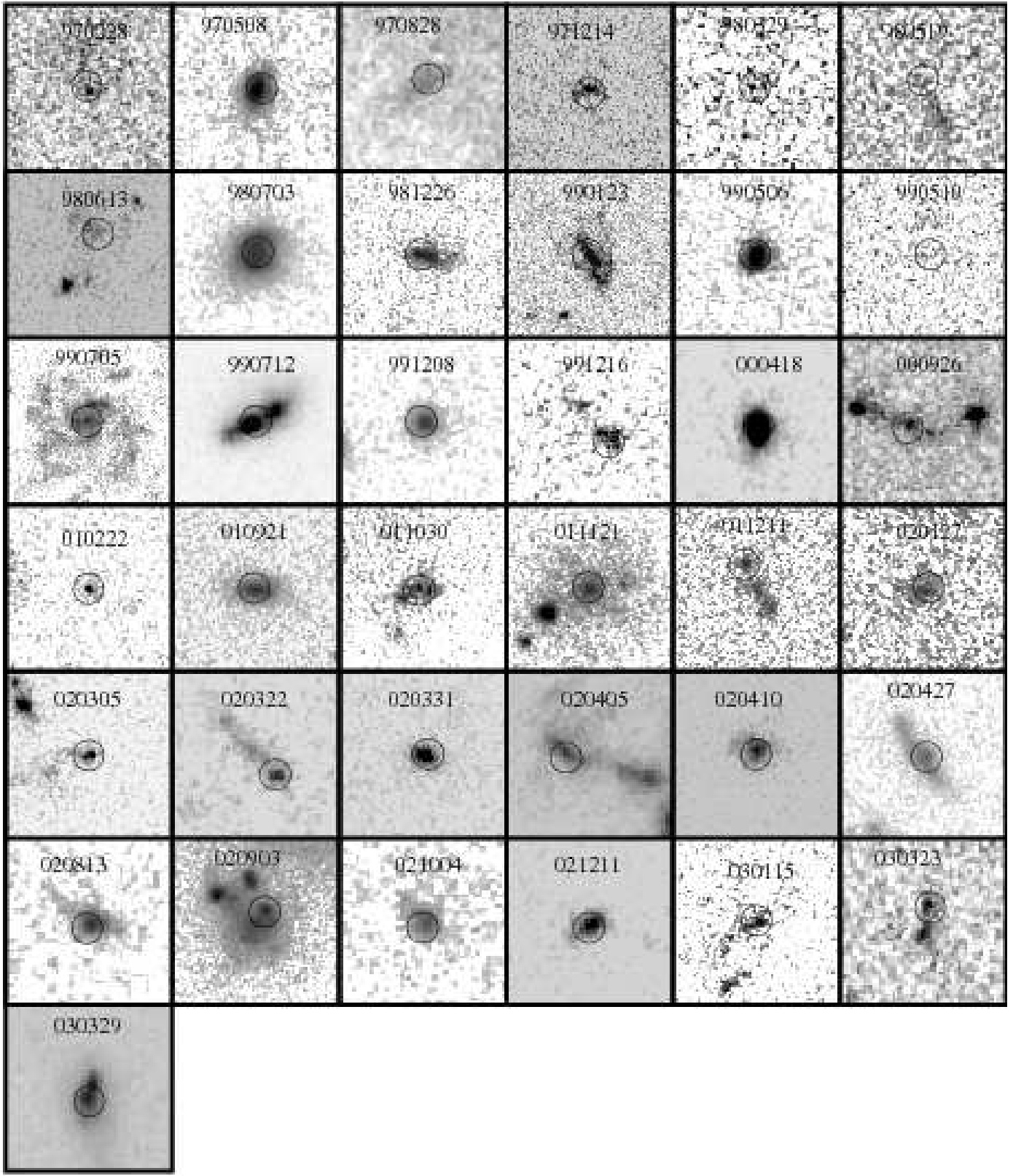}}
\end{center}
\vspace{-0.5cm}
\figcaption{Images of the GRB host galaxies used in this paper. The circled
region displays the initial center used in the CAS analysis at an
arbitrary size. }
\end{figure*}

The GRB hosts included in
our study are listed in Table~1 and are displayed in Figure~1.  Table~1 also
lists the instrument used to image each GRB host and the date of
the HST observation we use.  All of the hosts were observed months after
the GRB event, giving us a clean image of the host.   Many
hosts were imaged with STIS in the clear (50CCD) and longpass (LP)
filters, while others were imaged with WFPC2 and ACS in several filters. We 
used the 50CCD and F606W images for hosts imaged with WFPC2 or ACS.  
The images were bias-subtracted and 
flat-field corrected by the HST pipeline, and then drizzled
(Fruchter \& Hook 2002), resulting in images with half the original pixel 
scales of 0\farcs0254,
0\farcs05 and 0\farcs03 for STIS, WFPC2 and ACS, respectively. All images used
were taken sufficiently long after the burst, such that the contamination
from the early afterglow is negligible.  

\section{Size, Structural and Morphological Analyses}

One of the primary methods for comparing galaxies at different, and within
similar, redshifts, is to study their stellar light distributions 
(e.g., Conselice 2003) and sizes (e.g., Ferguson et al. 2004).
The manner in which the stellar light in galaxies is distributed can reveal 
the most salient features
of a galaxy's evolution and its characteristics, including: star formation,
the presence and history of interactions/mergers, and a galaxy's scale or
relative total mass.  Here we examine these features in a few
ways - first we look at gross light profiles by fitting exponential
and de Vaucouleurs r$^{1/4}$-law functions. Later we utilize model 
independent CAS (Conselice 2003)
parameters to compare the structures of the gamma-ray burst selected
galaxies with other galaxies at similar redshift to look for
evolution with redshift.  We also analyze the size distribution of
GRB hosts as a function of redshift.

\subsection{Profile Fitting}

\subsubsection{Method}

We fit exponential and de Vaucouleurs models to the surface
brightness profiles of eight bright galaxies, with $R < 24$ 
and with spectroscopic redshifts $z < 1.2$, from our sample of GRB
hosts imaged with STIS. The noise in the images of the fainter galaxies 
becomes too
large for a meaningful comparison between the two models.  We perform these 
fits to determine if GRB host galaxies have similar light distributions to 
present day ellipticals or disks.
Most elliptical galaxies and bulges of spiral galaxies are well-fit with a
de Vaucouleurs r$^{1/4}$-law profile: I(r)=I$_e$ exp
($-$7.669((r/R$_e$)$^{1/4}-$1)), where R$_e$ is the effective radius
corresponding to the isophote which contains half of the galaxy light,
and I$_e$ is the surface brightness at R$_e$. Disks of spiral
galaxies, however, are best fit with an exponential profile:
I(r)=I$_0$ exp ($-$r/R$_d$), where R$_d$ is the disk scale length, and
I$_0$ is the central surface brightness. 

\begin{figure*}
\begin{center}
\vspace{0cm}
\hspace{-1cm}
\rotatebox{0}{
\includegraphics[width=0.7\linewidth]{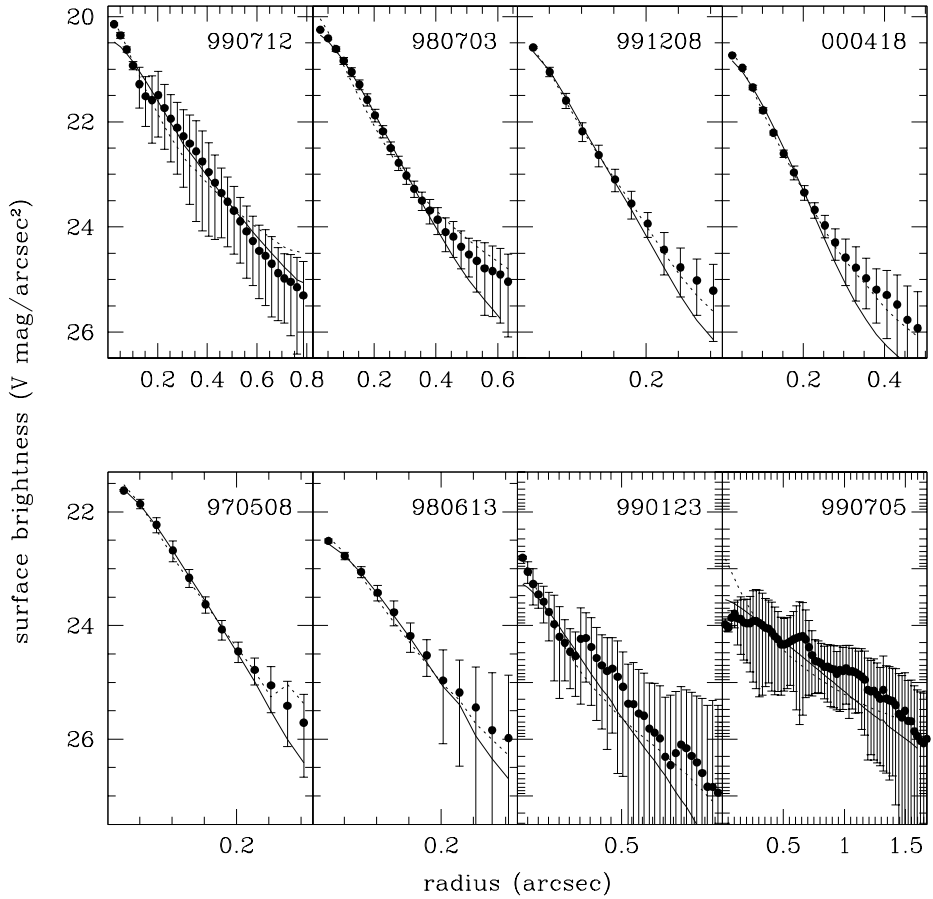}}
\end{center}
\vspace{-0.5cm}
\figcaption{The surface brightness profiles of eight bright host
  galaxies at $z < 1.2$ imaged with STIS.  The hosts are fit in 2 dimensions 
   with two models: an
  exponential disk, and an r$^{1/4}$ profile; both models are first
  convolved with the PSF before performing the fit. For the object
  image and the model images, we fit ellipses using the task {\it
    ellipse} in IRAF to obtain the surface brightness profiles shown
  above. The lines are not fits to the data points in these plots, but
  rather to surface brightness profiles of the best-fit model images.
  The solid line represents the exponential disk model, the dotted
  line the r$^{1/4}$, or de Vaucouleurs profile. See Table~2
   for the best fit parameters. }
\end{figure*}

Our fitting method is similar to the one used by Fruchter et al. (2000), 
who fitted exponential and r$^{1/4}$-law
profiles to the host of GRB\,970508.  For each image we first cut out
a small area around the host galaxy, from which we build a Poisson
error image. During each iteration in the fitting process, we create a
2-dimensional model image based on six parameters for both models.
These parameters are: the central or effective surface brightness
(I$_0$ or I$_e$, respectively), the disk scale length or effective
radius (R$_d$ or R$_e$, respectively), position angle (PA), axis
ratio, and x and y pixel position.  The model image is then convolved with
the point spread function (PSF) before fitting it to the observed
object image. To create a PSF image that approximates the real PSF of
the telescope and instrument as closely as possible, we drizzle
several model PSFs (as many as there are dithered individual object
images that were used to create the final object image).  Such a
``raw'' model PSF can be produced with Tiny Tim software (Krist 
1995)\footnote{see
  http://www.stsci.edu/software/tinytim/}.  This subsampled PSF image is
given a random offset in x- and y-position several times, just as the
raw object images are offset with respect to each other, typically by
10-20 pixels. These images are then rebinned to the original pixel
size, convolved with the charge diffusion kernel and drizzled to the
final PSF image using the same parameter values, such as the scale and
rotation, as in the drizzling run of the object image. As expected, the
resulting PSF is similar to the PSF of stars in the STIS images. After
convolution of the model image with this PSF, we used the IDL-amoeba, fits,
also known as the down-hill simplex method (Nelder \& Mead 1965; Press et 
al. 1992) to minimize 

$$\chi^2 = \left[\frac{\rm object -
    convolved(model, psf)}{\rm error}\right]^2.$$

\noindent The error for each fit parameter in Table 2 is estimated by changing
it in small increments from its best fit value while allowing all the other 
parameters to vary. The change for which the resulting $\chi^2$ increases by 
unity is taken to be the error in the parameter value.

\subsubsection{Results of the Profile Fitting}

The results of these fits for eight of the brightest host galaxies at $z < 1.2$
imaged with STIS are listed
in Table~2 and are graphically displayed in Figure~2.  This is not the first
study of GRB host galaxy structure, but most previous work contained
an analysis of only single objects, which generally agree with our
fits. For example, the host of GRB\,970508, 
(Fruchter et al. 2000) was best fit with an exponential disk with a 
scale length of 0\farcs046
$\pm$ 0\farcs006, and an ellipticity of 0.70 $\pm$ 0.07.    We find
almost the same result: the $\chi^2$ of the exponential disk profile
is clearly lower (0.57 vs. 0.85 for the r$^{1/4}$ profile). The axis 
ratio that we find
(0.39) corresponds to an ellipticity of 0.61, which is also consistent
with the result of Fruchter et al. (2000).  Although
Figure~2 seems to graphically show that the exponential and de Vaucouleur 
profiles
are equally well fit, the central part of the profile is 
better fit with the exponential. However, this is partially
misleading, as although  GRB~970508 is  
better fit by an exponential model, the  r$^{1/4}$ model cannot be 
ruled out by the data (with $\chi^{2}$=0.85).
Other examples can be seen from Table~2, and the model independent CAS
parameters (\S 3.2) are perhaps a better way to quantify and compare
the structures of faint high-z systems.

With this caveat in mind, we find that the exponential model 
provides the best fit for the hosts of GRB\,970508, GRB\,980703 and
GRB\,990712  (see Table~2 where the $\chi^{2}$ values for these
galaxies are listed for both fits). In the cases of 
GRB\,991208 and GRB\,000418, the data are
better fit with a de Vaucouleurs profile, although for these galaxies
the difference in $\chi^2$ between the two models is not very large.   An
early type interpretation however agrees with that found using the 
CAS methodology (Table~1; \S 3.2).
For the other three galaxies, both models fit the data nearly equally well 
(Table~2), and are more irregular/peculiar in appearance (\S 3.2).
It appears from this that the structures of GRB host galaxies are not all
identical, but have a diversity in profile shapes, which is confirmed by
visual impressions and the analysis in \S 3.2.

\subsection{CAS and Size Parameters}

\subsubsection{Background}

The CAS (concentration, asymmetry, clumpiness) parameters are useful
for characterizing the structural properties of galaxies using
model independent measurements. Unlike surface brightness
fits, CAS parameters can be measured for most of the GRB hosts, even the
fainter systems. Previously, Schade et al. (1995), 
Abraham et al. (1996), Conselice (1997), Bershady et al. (2000),
Conselice et al. (2000a,b), Conselice (2003) and Mobasher et al. (2004)
have shown that
galaxies can be roughly classified in three broad morphological classes: E/S0,
spiral, and peculiar/irregular/merging galaxies on the basis of their
central concentration and asymmetry.   We apply the methods of Conselice
(2003) to determine CAS parameters for our GRB host galaxies and
compare these with previous measurements of nearby and high redshift
galaxies.  The resulting values and types are listed in Table~1.

We also measure the sizes of the GRB host galaxies through the CAS method
which utilizes the Petrosian radius (e.g., Petrosian 1976; 
Bershady et al. 2000).   The
radius we measure for each galaxy is 1.5 times the radius where the inverse 
Petrosian index reaches $\eta = 0.2$. 

\subsubsection{Measurements}

The definition of the central
concentration ($C$) we used in this paper, following Bershady et al. (2000), 
is: $C$ = 5 log$_{10}$ (r$_{80}$/r$_{20}$),
where r$_{80}$ (r$_{20}$) is  the radius that contains 80\%
(20\%) of the galaxy light. A galaxy with a steep profile, such as an
elliptical, will show a relatively large value for the concentration
parameter ($C > 3.5$), while galaxies with a more shallow light profile, such 
as spiral and irregular galaxies, will have lower $C$ values. 

The asymmetry parameter ($A$) is determined by rotating a galaxy by 
180$\degr$ from its center, and
subtracting the rotated image from the original image.  A perfectly
symmetric galaxy will show no residuals in the difference image, while
a galaxy with asymmetric features such as bright star-forming regions,
or an interacting galaxy, will have large residuals. The absolute
value of the pixels in the difference image, normalized by the pixel
values in the original image, and corrected for the background,
gives a measure of the asymmetry parameter
(Conselice et al. 2000a, Conselice 2003),

$$A= {\rm min} \frac{\Sigma |(I_o - I_{\phi})|}{\Sigma |I_{o}|} - {\rm min} \frac{\Sigma |(B_o - B_{\phi})|}{\Sigma |I_{o}|}.$$

\noindent where $I_o$
and $I_{\phi}$ are the pixel intensities in the original and rotated
image, respectively, and the corresponding $B$ values
represent the blank field (background) 
region used to account for the intrinsic asymmetry due to the background 
noise.
Determination of the center of rotation is
important, as is the radius in which the parameter is measured (see
Conselice et al. 2000a). The asymmetry routine computes the asymmetries on a 
grid of
centers around an initial center value until a minimum asymmetry is found
(within 5 pixels).  We always used $\phi = 90$\deg for the asymmetry 
computations.
The clumpiness index ($S$), which is defined in a similar way,
 is too difficult to measure
for most of the GRB hosts due to their faintness, and we 
do not consider this parameter in our analysis.  To calculate the concentration
and asymmetry parameters we place the initial center for our analysis on
the location shown by the circles on Figure~1.  The code measures
the Petrosian radii and then measures the parameters within this determined
radius
(Conselice 2003).  For the irregular and merger systems shown in 
Figure~1 the other `parts' of the host are included if they
are within this radius.  This is the same method used on the Hubble Deep
Field data sets, and thus allows for a fair comparison.

\subsubsection{Biases}

Although this classification method is very promising, there are some
caveats. One  is that when observing high-redshift galaxies,
one has to be careful when comparing values with those of local
galaxies because of bandshifting effects: the rest-wavelength 
probed is bluer and therefore galaxies may appear more
asymmetric due to the dominant patchy younger stellar 
populations (e.g., Windhorst et al. 2002). However,
from a multi-wavelength study of
a small sample of nearby starburst galaxies
Conselice et al. (2000c) find that starburst galaxy morphology changes
little from the visible to the ultra-violet (UV) wavelength regimes,
suggesting that the inferred morphology of starburst galaxies at
high redshift is likely similar to their local classification.
Investigating the wavelength-dependent morphology of nearby
galaxies, Kuchinski et al. (2000) and Windhorst et al. (2002)
also find that the change in apparent
morphology from the visible to UV is dramatic for early-type spirals
with prominent bulges, but modest for late-type spirals and
irregulars. These results suggest that the change in inferred
morphology with increasing redshift for GRB host galaxies, which on
the basis of their star-formation rate are thought to be actively
star-forming galaxies, is unlikely to be very large.

To understand quantitatively the systematics in measuring parameters at high redshifts, 
we carried out a series of simulations of nearby galaxies to determine how
CAS and size parameters change with distance.   These
simulations include surface brightness dimming, reduced resolution, and
increased amounts of added noise.  We degrade the images
of nearby galaxies, to be at the
GRB redshifts used in this paper, with the same
exposure time used to image the GRB hosts.  We used these new images
to determine how changes in
redshift result in different CA measurements.  We carried out these simulations
using 82 nearby normal galaxies of all Hubble types, the same used in the 
simulations discussed in Conselice (2003). We find that the concentration 
and asymmetry values decline, on average,
by  $\delta C = 0.25\pm 0.43$ and $\delta A = 0.08\pm 0.1$ due to
redshift effects.  These values change only slightly at higher
and lower redshifts than the median $z \sim 1.2$ redshift. 
Furthermore, we find that on average, a galaxy will appear 25\% smaller
in the STIS/ACS/WFPC2 host imaging due to surface brightness dimming.  
 Unless specified we however do not generally apply these 
corrections to the CAS and radii values.

\section{Results}

\subsection{GRB Host Comparison with Field Galaxies}

\begin{inlinefigure}
\begin{center}
\vspace{2cm}
\hspace{-1cm}
\rotatebox{0}{
\resizebox{\textwidth}{!}{\includegraphics[bb = 45 45 550 550]{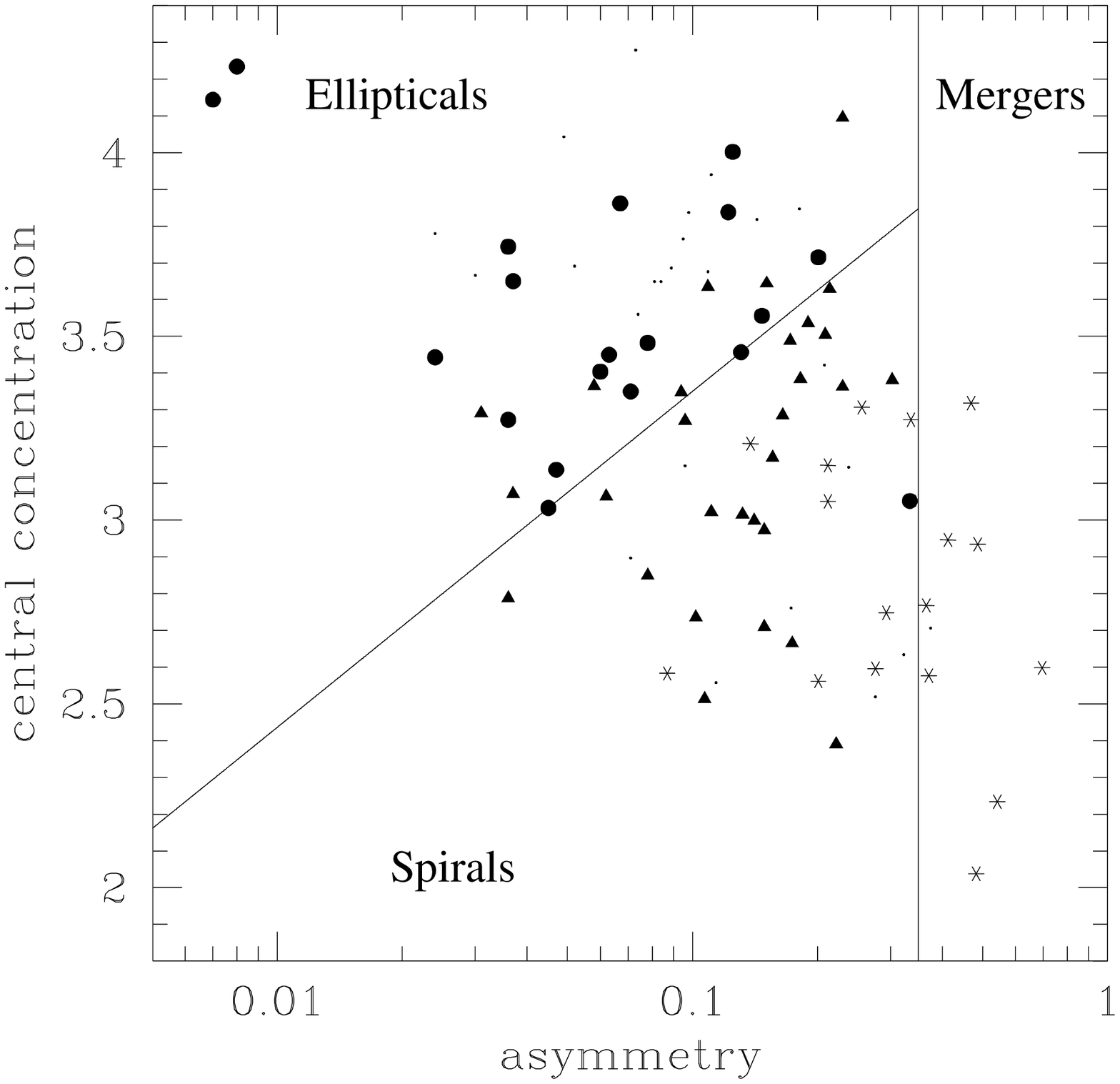}}}
\end{center}
\vspace{-1cm}
\figcaption{The concentration versus asymmetry diagram for galaxies brighter
  than $I=24$ in the HDF.   The different symbols correspond to the broad
  Hubble type bins into which the galaxies were visually classified: 
  ellipticals are
  shown as filled circles, spirals as filled triangles, peculiar
  galaxies as stars and unclassified galaxies with $I < 24$ as dots. 
  Although there is some overlap, three different
  regions can be distinguished, which are separated by the solid
  lines. Local galaxies with an asymmetry larger than 0.35 tend to be
  mergers. The line separating the elliptical from the spiral region
  is a ``best estimate'' by eye. }
\end{inlinefigure}

We used three galaxy samples to compare the measured
asymmetries and central concentrations of the GRB host galaxies.  The
first one is a BVI summed image of the Hubble Deep Field North (HDF-N)
(Williams et al. 1996), for which a visual classification is
available for galaxies  between $I=21$ and $I=25$
(van den Bergh et al. 1996). We used a BVI
combined image of the HDF-N to match the wide wavelength range of the 
GRB host observations, although using individual bands gives the same results.
We have performed two CAS measurement runs on the HDF: one with the full 
depth, in order to compare the concentration-asymmetry (CA) values that we 
find with the visual classification of HDF galaxies (Figure~3), and 
one where we have 
degraded the HDF-N image to match the typical noise level of the images
containing the GRB hosts (Figure~4). We did not rebin the images to match the
pixel-size of the STIS images (0\farcs0254) with that of the HDF
(0\farcs04). We add noise to this sum to scale it to the
typical exposure time of the host galaxy images (exposure time ratio
is roughly 25), and calculate the CA values for the detected galaxies
(crosses in Figure~4).   For a second comparison sample we have measured 
CA values for all the
galaxies that are detected in the GRB host images.  This second sample
allows for a direct comparison of the morphology of the hosts with that of
an unbiased sample of galaxies that were imaged with the same pixel
size, depth, and filter.  A third comparison is Hubble Deep Field galaxy 
rest-frame CAS parameters with redshifts from Conselice et al. (2005).

We used the SExtractor code (Bertin \& Arnouts 1996) to detect
objects on both the GRB and HDF images, and we used these positions as input 
for the CAS code to calculate asymmetry and concentration indices.  For the 
object detection the
images are first smoothed with a Gaussian function, after which we
included objects in our catalog that have at least 3 contiguous pixels
that are each 2$\sigma$ above the sky background. For the full depth
HDF we relaxed this condition to 10$\sigma$, as we are not interested
in the very faintest objects.  We used SExtractor's star-galaxy
separator stellaricity index to eliminate stars, by removing
objects with values $> 0.8$. 
The remaining objects are checked for saturation, and put into the CAS 
routines.
SExtractor sometimes finds several ``centers'' (peaks) for the
same galaxy.  We investigated if there is another peak
present within a radius of 0\farcs5 for these systems. If so, we 
choose the peak with
the lowest value for the asymmetry, and discard the other peak.  We
perform the same exercise for the host galaxies. As an example, for
GRB\,991208, SExtractor 
picks up a peak eastward of the presumed host.  Centering at this
peak, the asymmetry routine measures a much larger asymmetry, and is therefore
discarded.  We also removed all foreground stars and background/foreground
galaxies near each host galaxy before
measuring $A$ and $C$ values.  

The results of these analyses are shown in Figure~3 where 
we plot a C-A diagram for the
galaxies brighter than roughly $I$=24 in the HDF-N. As shown by
 Conselice (2003) and Conselice et al. (2003), galaxies of different 
Hubble types are
located in different parts of this diagram. These broad bins can be
roughly separated by the shown solid lines. The vertical line at an
asymmetry value of $A = 0.35$ corresponds to the threshold above which nearly
all galaxies are mergers (see Conselice et al. 2000b; 
Conselice 2003; Hernandez-Toledo et al. 2005). In Figure~4 we show the 
CA values for 37 host galaxies compared with all the other
galaxies detected in the same images (solid dots).

Examining Figures~3 and 4 and applying the concentration and asymmetries
corrections discussed in \S 3.2.3 places a few
more GRB hosts in the region of major mergers than what their measured
location suggests.  This reveals that GRB hosts have a slightly
higher than average merger rate at $z < 1.2$ compared with photometrically
selected field samples (e.g., Conselice et al. 2003a;
Lin et al. 2004; Bundy et al. 2004), although mergers cannot account
for all GRB hosts.

\subsection{GRB Host Galaxy Structures}

It appears from  Figures 1-4, and the discussion in \S 3, that GRB host 
galaxies arise in all types of field galaxies, not just irregular or peculiar
starbursts.  For the sample of eight bright hosts which we fit exponential and
de Vaucouleur profiles, we find that the GRB
hosts are not uniformly either spiral galaxy-like
or elliptical-like, but are a mix of all types.   The scale lengths of the 
hosts also range from very small, 
0.14 kpc in the case of GRB\,991208, to a value that is comparable to the
Galactic scale length: 4.7 kpc for the host of GRB\,990705.  This effectively
also spans the range in the sizes of field galaxies at these redshifts 
(\S 4.3).

We used the value of the
central concentration and asymmetry of a galaxy as a rough
indication for its morphology, and as a means of quantifying the differences
between GRB hosts and other field galaxies. 
Both the HDF and field galaxy comparison samples occupy roughly
the same regions in the C-A space as the GRB hosts (Figures 3 \& 4), 
which suggests that 
GRB host galaxies are typical field galaxies, and are not exclusively
mergers or peculiars, such as sub-mm galaxies and bright
Lyman-break galaxies (Conselice et al. 2003a,b).
Many hosts have C-A values in the spiral region close
to the ``border'' with the local merger population. For example, the
host of GRB\,990705 is clearly a ``grand-design'' late-type spiral
(Le Floc'h et al. 2002), in agreement with its position in the C-A diagram.

\begin{inlinefigure}
\begin{center}
\vspace{3cm}
\hspace{-1.5cm}
\rotatebox{0}{
\resizebox{\textwidth}{!}{\includegraphics[bb = 45 45 550 550]{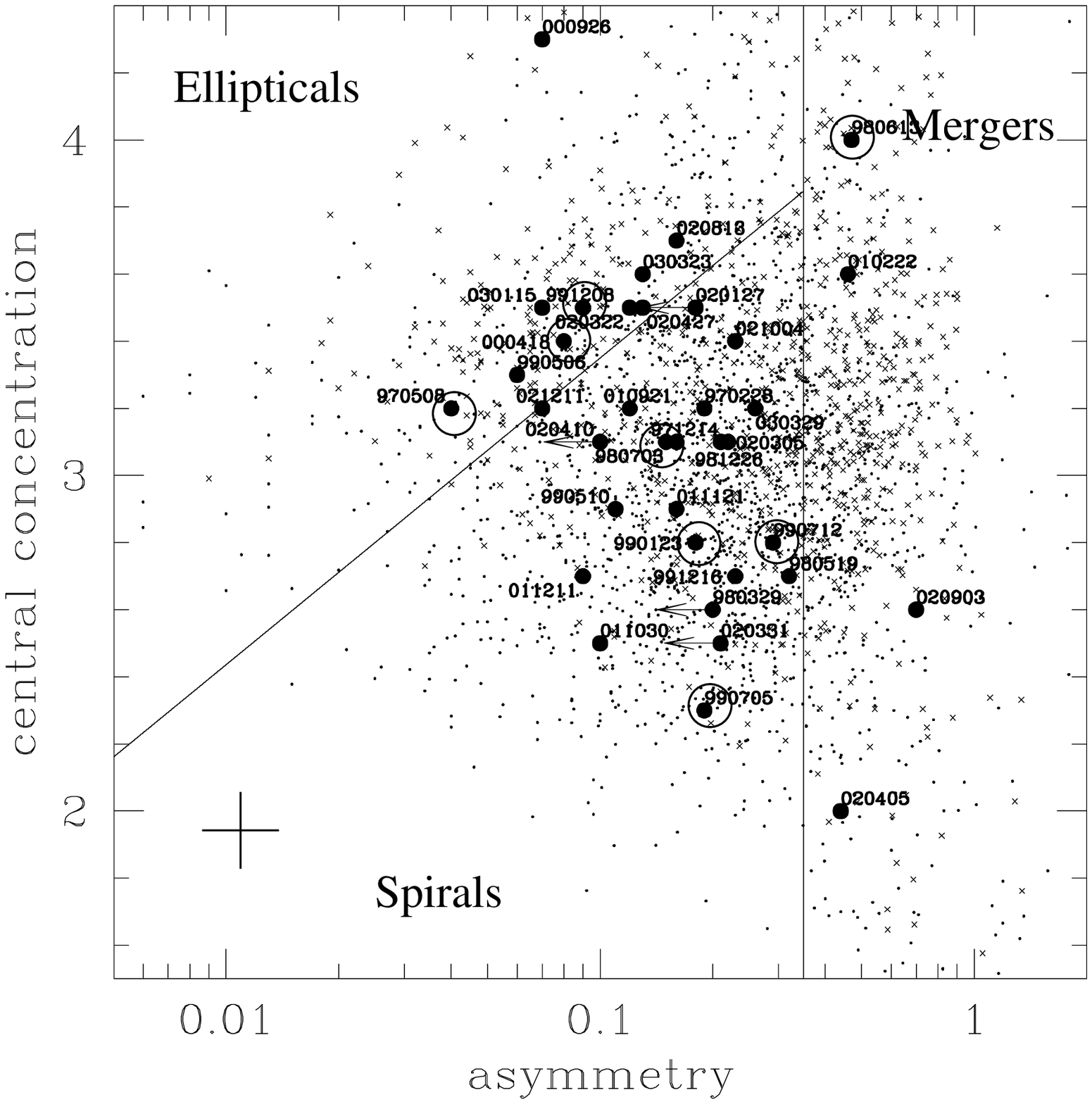}}}
\end{center}
\vspace{-1cm}
\figcaption{The concentration versus asymmetry diagram for our sample of 
  host galaxies,
  together with all other galaxies that are present in the GRB host
  galaxy images (solid dots), and galaxies in a BVI sum of the images
  of the HDF (crosses), with added noise to mimic the typical exposure
  time of the host-galaxy images. The systems which are circled are those
  for which we performed surface brightness fits (\S 3.1).   
  The host GRB~970828 is not plotted here 
  due to its negative asymmetry value. }
\end{inlinefigure}

Several of the GRB hosts studied in
this paper has been examined previously.  Odewahn et al. (1998) analyzed
the morphology of the host of GRB~971214 using
surface brightness fits and apparent morphology.  Odewahn conclude, as
we do, that this host has
a compact but irregular structure.  
Similarly, the host of GRB~980703, studied by Holland et al. (2001)
with STIS on HST, is found to be a late-type disk with asymmetric
structures, similar to what we find. The host of GRB~980613
was previously studied by Hjorth et al. (2002), who find an asymmetric
and chaotic structure, while Gorosabel et al. (2003) argue that
the host of GRB~000418 is compact and smooth, both in agreement with
our results (Table~1).   In total, we find that the
hosts of GRB\,980613, GRB\,010222,
GRB\,020405 and GRB\,020903 are all located in the peculiar section/merger
area of the C-A diagram.   For GRB\,980613, GRB\,020405 and GRB\,020903
this is not unexpected, since these all appear very peculiar.  For
GRB\,010222 this is less obvious, although for this galaxy a sub-mm
flux is measured that is consistent with the host being an intense starburst
galaxy (Frail et al. 2002). 

Interestingly, the CA values for twelve hosts are consistent with
an early type morphology.   Two of these, GRB\,000418 and GRB\,010222, have
a very high star-formation rate  inferred
from sub-mm observations (Frail et al. 2002;
Berger et al. 2003), which might
seem contradictory to their elliptical appearances.   Low luminosity
ellipticals at $z \sim 1$ are however generally blue with star formation 
throughout their structures (Stanford et al. 2004).  
Furthermore, Arp 220-like galaxies, probable 
ellipticals in formation, have light
profiles with a r$^{1/4}$ form, and contain high light concentrations.  
Assuming that GRBs are related to the deaths of massive
stars, and that nuclear starbursts mimic an elliptical appearance, the
concentrated hosts may contain nuclear
starbursts. In fact, the projected afterglow position of
GRB\,970508 is so close to the center of the galaxy, that
Fruchter et al. (2000) suggested that the burst may originate
from such a nuclear starburst forming into an early-type galaxy. 

In conclusion, from both surface brightness profile fitting and through
measuring the central concentration and asymmetry of a sample of GRB
host galaxies, we find that GRB hosts do not fit into one clear
single morphological class of galaxy. In the concentration-asymmetry
diagram, most GRB hosts are consistent with spirals or
irregular galaxies, although galaxies consistent with being mergers and
early types are also found.   This likely reveals that massive 
star formation is not produced in one single method, but in several 
different ways, such as major and minor
mergers, and the accretion of gas from the intergalactic medium.

\subsection{Evolution of Host Galaxy Sizes}

We measure the sizes of our GRB host galaxy sample utilizing the Petrosian
radius (Bershady et al. 2000; Graham et al. 2005), which for our
purposes is defined as 1.5 $\times$ r($\eta$) = 0.2 (Table~1).  We also
compute the identical radius for galaxies within the HDF 
(Conselice et al. 2005).  The GRB host galaxy sizes have been corrected
by adding an additional amount to the measured sizes, based on the 
redshift of the host and the results of our simulations (see \S 3.2.3).  This
procedure adds on average an additional 25\% to the sizes of the GRB hosts.  
We also perform a similar, but smaller,
correction to the measured sizes of the HDF galaxies (see 
Conselice et al. 2005 for details).  

As can been seen from Table~1, there are many $z < 1$ GRB host
galaxies with very small Petrosian radii $< 5$ kpc, although there are a few
exceptions (GRBs~970828, 990705, 011121, 020405). The average Petrosian
radius of the GRB hosts at $z < 1.2$ is 6.7$\pm5.2$ kpc (quoted
errors are the standard error of the mean), after removing
systems which are merging based on the asymmetry index.   This is
smaller than the average Petrosian radius of 12.0$\pm$8.5 kpc for
field galaxies at $z < 1.2$.  The average GRB host size at $z < 1.2$
reduces even further to 4.5$\pm1.5$ kpc if we remove the GRB~990705 host,
which is a large spiral system at $z = 0.84$.
On the other hand, the GRB host galaxy sizes at $z > 1.2$ tend to be roughly
the same as the average size of field galaxies at these redshifts.  The average
Petrosian radius for GRB hosts at $z > 1.2$ is 6.8$\pm$4.0 kpc, compared
with an average field galaxy Petrosian radius of 7.1$\pm$2.0 kpc.  
Thus, it appears that the sizes of GRB hosts may not
change significantly, but that the general field population does, such
that GRB hosts are less representative of the galaxy population at lower
redshift, than at higher redshift.

\subsection{Evolution of GRB Host Galaxy Structure}

We have a large enough sample of host galaxies to
study the evolution of CAS parameters as a function of redshift.
We can also compare GRB host properties to the properties of field
galaxies in the HDF at similar redshifts.  As discussed in \S 4.1
we find little difference between the concentration ($C$) and
asymmetry parameters ($A$) for the GRB hosts compared with HDF
galaxies in general.  This extends when we compare HDF field galaxies
at low redshift to the GRB hosts found at $z < 1.2$.
For this
comparison we divide our GRB sample into two redshift ranges at $z < 1.2$
and $z > 1.2$ which contains roughly the same number of hosts.  
We find that the hosts at $z > 1.2$ are different from the general $z > 1.2$
galaxy population in terms of their light concentrations.

\begin{figure*}
\begin{center}
\vspace{0cm}
\hspace{-1cm}
\rotatebox{0}{
\includegraphics[width=1\linewidth]{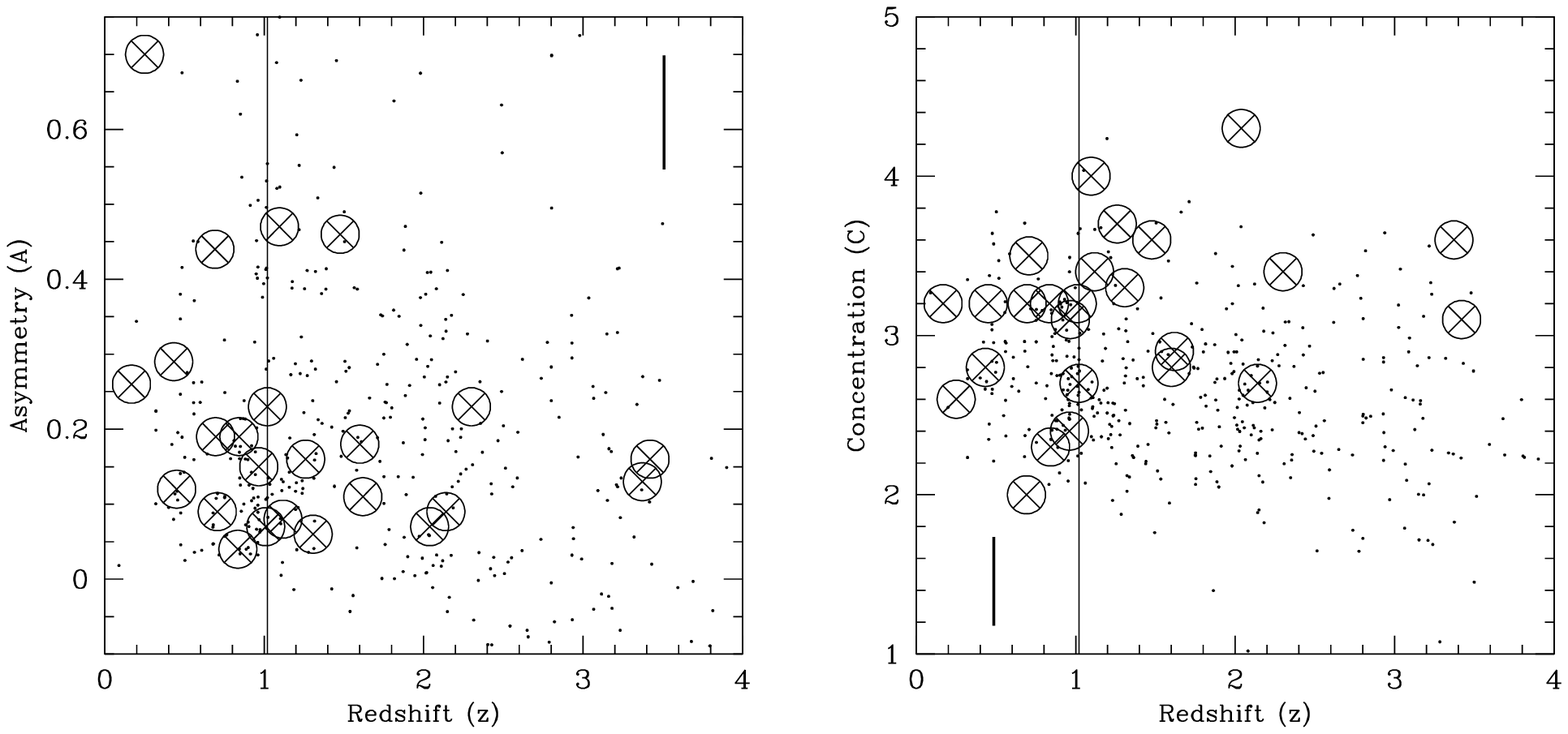}}
\end{center}
\vspace{-0.5cm}
\figcaption{The evolution of CAS parameters for GRB hosts as a function of
redshift.  The circled crosses represent the concentration and asymmetry
parameters for the hosts, while the tiny dots are field galaxies in the HDF
with redshifts (see Conselice et al. 2005).    
The vertical line divides the sample into high
and low redshift at $z = 1$. (see text).  No corrections to these values
have been applied, although implementing the corrections discussed in
\S 3.2.3 do not change the results. }
\end{figure*}

Concentration and asymmetry values are plotted as function of redshift for
the GRB hosts (circled crosses) and for HDF galaxies (tiny dots) on Figure~5.  
From this figure, it can be seen that the
GRB hosts occupy the high-C range at $z > 1$. 
A two-dimensional K-S test shows that there is only a 2.5 $\times 10^{-3}$
chance that the $C$ values of the GRB hosts, and those of HDF field
galaxies, arise from the same distribution, and our sample size is large
enough to make this result statistically meaningful. This probability
becomes higher when we apply the corrections discussed in \S 3.2.3. 
It thus appears that
at $z > 1$ the majority of detected GRB hosts are highly concentrated
galaxies, suggesting that they are forming into
elliptical galaxies, or are blue compact galaxies.   These highly
concentrated galaxies are also fairly large, with a few among the largest
GRB hosts known.   The concentration index also broadly correlates 
with stellar mass such that higher concentrated objects contain a higher
stellar mass (Conselice et al. 2005), 
suggesting that these galaxies could be more massive than average field
galaxies at $z > 1$.  

There are several things which could be biasing this result. The first
is that not all GRB hosts have redshifts and some are therefore not included
in our analysis. However, as Table~1 shows, many of the hosts without
redshifts have high concentration indices, with GRBs~020127, 020322, 020427, 
030115 all with $C > 3.5$, and because these GRB hosts are faint, they are more
likely at higher redshifts.   There is also no correlation between the
magnitude of the host galaxies and the concentration index, thus it is
unlikely that the faint hosts we are not including are predominately of lower 
concentration.  Thus,
not including these few systems is not likely to bias our results. Our
results are also unlikely to be the result of a morphological k-correction 
bias, as high redshift galaxies become
less concentrated when viewed in the rest-frame UV, 
and starbursting galaxies look very similar in the UV and optical 
(Conselice et al. 2000c; Windhorst et al. 2003; Papovich et al. 2003). The
structures of GRB hosts are therefore not expected to change much with 
wavelength,
as they are for the most part starbursting systems (see also 
Christensen, Hjorth \& Gorosabel 2004). 

This is another indication that GRB hosts are involved in star formation, and
importantly trace out galaxies undergoing rapid star formation.  The logic
behind this is based on a comparison to HDF galaxies, as follows.
It appears that at high redshift ($z > 1$) the most
concentrated galaxies account for a significant fraction
of all ongoing star formation, while
at lower redshift the star formation density is occupied by
lower mass systems (Conselice et al. 2005).   This is also seen in
the GRB hosts, under the  assumption that GRB hosts are star forming
galaxies.
This is likely a signature of the down-sizing of galaxy formation seen in 
other aspects as well (e.g., Cowie et al. 1996; Bundy et al. 2005).  

This evolution in 
starburst behavior can be seen by examining
the stellar mass and luminosity attached to galaxies in the HDF-N with
concentrations as large as the ones found for the GRB hosts.
In the HDF between $1 < z < 4$, only 9$\pm$2\% of the stellar mass is 
attached
to galaxies with $C$ values as high as the GRB hosts (see Conselice et al.
2005).  The rest-frame B-band luminosity
fraction for galaxies with concentrations this high is 17$\pm$3\%.   Thus,
the ratio of luminosity to stellar mass for field
galaxies with concentrations as large as the GRB hosts at these
redshifts is about a factor of two, suggesting that highly concentrated
galaxies are undergoing a disproportionate amount of star formation for their
stellar mass, and GRB hosts at $z > 2$ are among these systems.   There
is also a suggestion that other starbursting populations at high
redshift have similar CAS, particular high concentration, values (Chapman
et al. 2003; Conselice et al. 2003b).   We still
only have a limited number of GRB host images however, and future observations
are required before we can definitely place constraints on the entire GRB
population. Other properties such as direct measurements of stellar masses 
are needed 
to make definite comparisons to field galaxies selected by other methods
at $z > 2$.  The {\em Swift} satellite, combined with deep HST imaging or
ground based adaptive optics, should revolutionize our knowledge of GRB
host galaxies in the coming years, and potentially extend these
results to even high redshifts.

\section{Discussion}

Our results can be summarized by the following: first, the morphological 
distribution of GRB hosts includes all galaxy morphological types, including
spirals, ellipticals, irregulars and peculiars/mergers.  There is not
a single morphological type where GRBs are likely to trigger.   Second, we
claim that the structures of GRB hosts changes with time,
such that hosts found at $z > 1$ are more concentrated.  Because
highly concentrated galaxies in both the nearby and distant universe
are typically early-types, or early-types in formation, and tend to have a 
larger stellar mass than less concentrated galaxies, we conclude that the 
higher redshift GRB hosts
are possibly more massive than a typical $z > 1$ field
galaxy. Stated another way, galaxies
in which GRBs are found change with redshift, such that at the
highest redshifts, they may not be small low mass systems,
as they tend to be at lower redshifts.  
We however are not claiming that GRB hosts at $z > 1$ 
are situation within the most massive starbursting systems, such as 
sub-mm sources, but that they are among the more massive average
field galaxies at these redshifts.  Our morphological and size measurements for
GRB hosts at $z < 1$ is consistent with the interpretation of these
systems as mostly small, lower mass galaxies.

There has been some previous work on the higher redshift GRBs studied in
this paper that may contradict the interpretation of these systems
as $> {\rm L}_{*}$ galaxies. It is fairly clear
that GRB hosts are not all dusty starbursts, as revealed through a lack
of sub-mm detections for most known hosts (Tanvir et al. 2004; Smith et al
2005).  However,
some GRB hosts, such as GRB~010222, are detected in the sub-mm, with a large
inferred star formation rate and a large bolometric flux (Frail et al.
2002). There are also examples of radio luminous hosts, such as GRB~980703, 
which have total infrared luminosities of $> 10^{12}$ L$_{\odot}$ and
star formation rates approaching 1000 M$_{\odot}$ yr${-1}$ (Berger
et al. 2001).  Furthermore, Chary et al. (2002) examined the spectral energy
distributions of 12 GRB hosts and concluded that three have
infrared luminosities comparable to infrared luminous galaxies.

There is also some evidence that GRBs are situated in regions of low
dust extinction. For example, Vreeswijk et al. (2004) studied 
the properties of the gas
in GRB~030323 and found a low metallicity environment, and 
a high column density
of neutral hydrogen in the region around the burst.  Hjorth
et al. (2003b) find a similar pattern for GRB~020124 located in 
a damped lyman-alpha system at $z = 3.2$.  Also, GRB hosts are
more often Lyman-alpha emitters than Lyman-break galaxies (Fynbo et al.
2003).  These results tend to imply
that GRB host galaxies are largely dust-free systems, yet GRBs may be situated
in low-metallicity regions of distant galaxies, or the intense energy
ejecta from GRBs destroys any nearby dust (e.g., Galama et al. 2003).
Also, Savaglio et al. (2004) find that some GRB hosts
contain evidence for large dust depletion, thus it is not clear
if GRB hosts are all dust free systems.

Finally, several studies have argued that GRB host galaxies
are sub-massive and blue systems (e.g., Le Floc'h et al. 2003) based
on near infrared imaging.  The Le Floc'h et al. (2003) study, and others
(e.g., Chary et al. 2002) are based on small samples of 
$\sim 10$ GRB hosts, most of which are at $z < 1$. Our results agree with
the conclusion from these studies, as we find that GRB hosts 
at $z < 1$ tend to be taken from the smallest field galaxies at $z \sim 1$.  
However, these studies do not contain a significant number of sources
at the highest redshifts, those at $z > 1$, which tend to have higher
luminosities. In fact, the highest redshift source in Le Floc'h et al.
(2003), GRB~971214, at 
$z = 3.42$, is the most luminous in their sample with an absolute magnitude of 
M$_{\rm K} = -24.45$.   

Our results indicate that the hosts at $z < 1$ are indeed small star forming
galaxies,
with a range of morphologies.  At $z > 1$, the situation is different, such
that the hosts are more concentrated than the average field galaxy at
similar redshifts at a confidence of $> 99.5$\%.  The average size
for hosts at $z > 1.2$ is 6.8$\pm$4.0 kpc, similar to within the errors
of the average Petrosian radius for field galaxies at similar redshifts, and
for galaxies with similar concentration indices.
At lower redshifts, the concentration index for hosts becomes lower, and the
sizes become smaller, relative to the average field galaxy.  This implies that
the sizes of GRB host galaxies do not change significantly with time, while 
the general
field galaxy population becomes larger.  
On the other hand, the higher concentration index
could either imply that these are progenitors of massive galaxies, or
that they are blue compact galaxies, which also tend to have similar
high concentration values (Jangren et al. 2005 in prep).  In either case, the
type of galaxy in which GRBs are found changes with redshift into the
population that dominates the star formation.

\section{Summary}

In this paper we study the structural parameters of 37 Gamma-ray bursts (GRB) 
host galaxies imaged
by the {\it Hubble Space Telescope}.   We used two methods to characterize the
structures of these galaxies. For  the brightest systems at 
$z < 1$ we fitted exponential and de Vaucouleur r$^{1/4}$ 
profiles, determining that GRB hosts
are not uniform, as some are better fit with the exponential than the de 
Vaucouleur profile, and vice versa.  We also measured the concentration and
asymmetry parameters for all 37 host galaxies, finding that the hosts are
not selected from one type of galaxy, but span a range from highly
concentrated systems resembling ellipticals in formation, disk like galaxies,
and systems undergoing merging.  

After dividing the sample into two redshift ranges we find that the higher
redshift ($z > 1$) GRB hosts have a higher concentration index than field galaxies
at similar redshifts, at a significance level $> 99.5$\%, yet they have similar sizes.    
Using the Hubble Deep Field North we also show that systems with concentration values 
this high at $z > 1$ are undergoing a disproportionate amount of star formation.  On the
other hand, lower
redshift systems at $z < 1$ are found to span all morphological types, and have 
sizes smaller than the average field population within the same redshift range.  This
is perhaps an indication that the nature of GRB hosts changes as a function of
redshift.  While lower redshift GRB hosts are smaller and perhaps
lower mass bluer galaxies (e.g., Le Floc'h et al. 2003), higher redshift
hosts are more concentrated and have typical field galaxy sizes.  
This is likely due to the fact that 
galaxies in which star formation occurs changes with redshift, with
the most massive systems undergoing the most formation at higher redshifts
(e.g., Cowie et al. 1996; Conselice et al. 2003; Heavens et al. 2004; Bundy
et al. 2005).  This
further suggests that GRBs are potentially one of the best ways to locate the
first galaxies forming at $z > 7$ in the densest areas of the universe.

\acknowledgements 

It is a pleasure to thank Sacha Hony for his expertise and help, and
Rachel Gibbons and Joris Gerssen for their help in fitting surface
brightness profiles. We also thank  Lex Kaper, Jens Hjorth and
Isabel Salamanca for valuable input. This study is partly based on
observations that were made as part of the Survey of the Host Galaxies
of Gamma-Ray Bursts. CJC acknowledges support from an NSF Astronomy and
Astrophysics Fellowship.  We thank the referees for their comments on
this paper.

\begin{figure*}
\begin{center}
\vspace{0cm}
\hspace{-1cm}
\rotatebox{0}{
\includegraphics[width=1\linewidth]{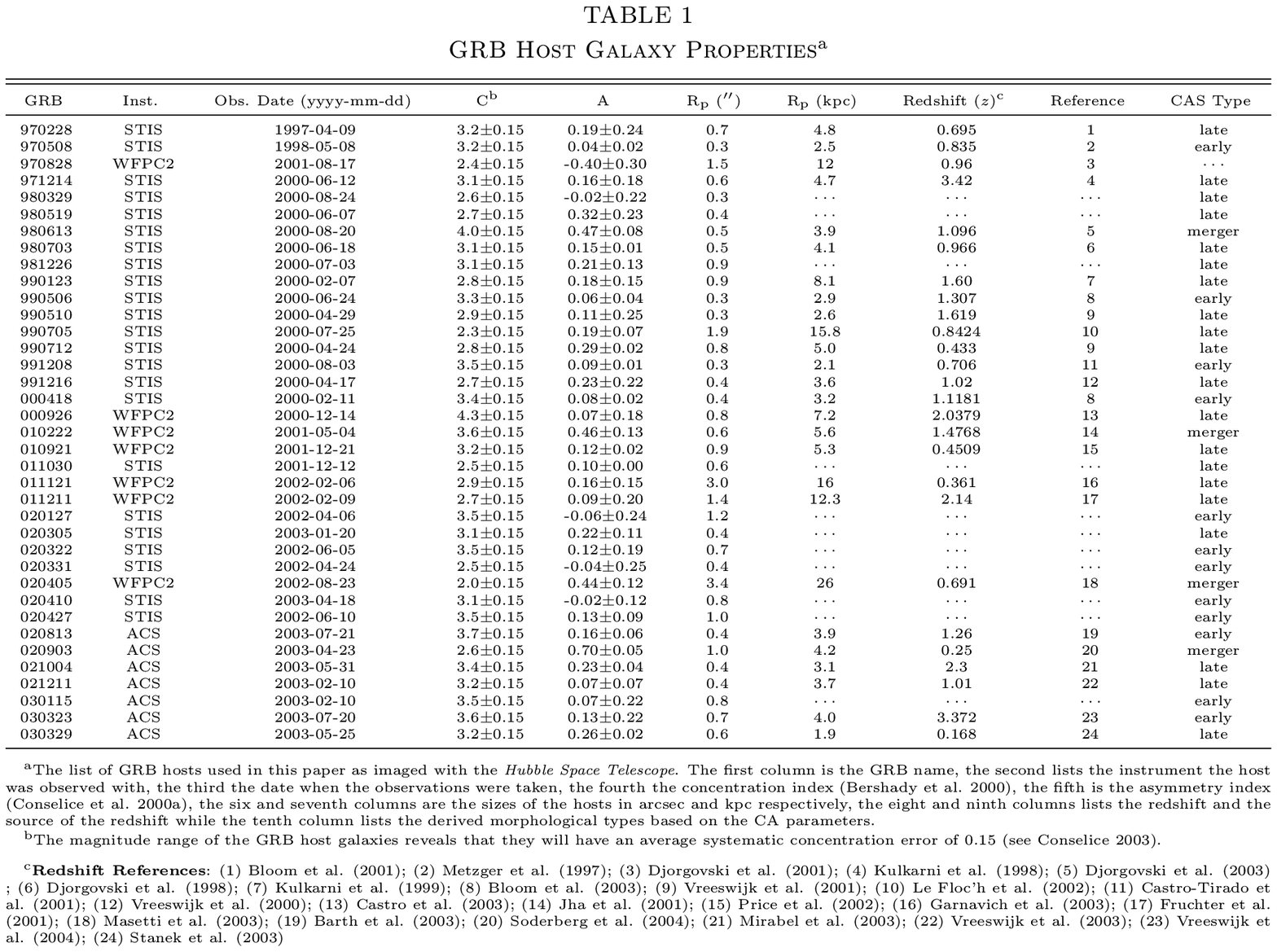}}
\end{center}
\vspace{-0.5cm}
\end{figure*}

\begin{figure*}
\begin{center}
\vspace{0cm}
\hspace{-1cm}
\rotatebox{0}{
\includegraphics[width=1\linewidth]{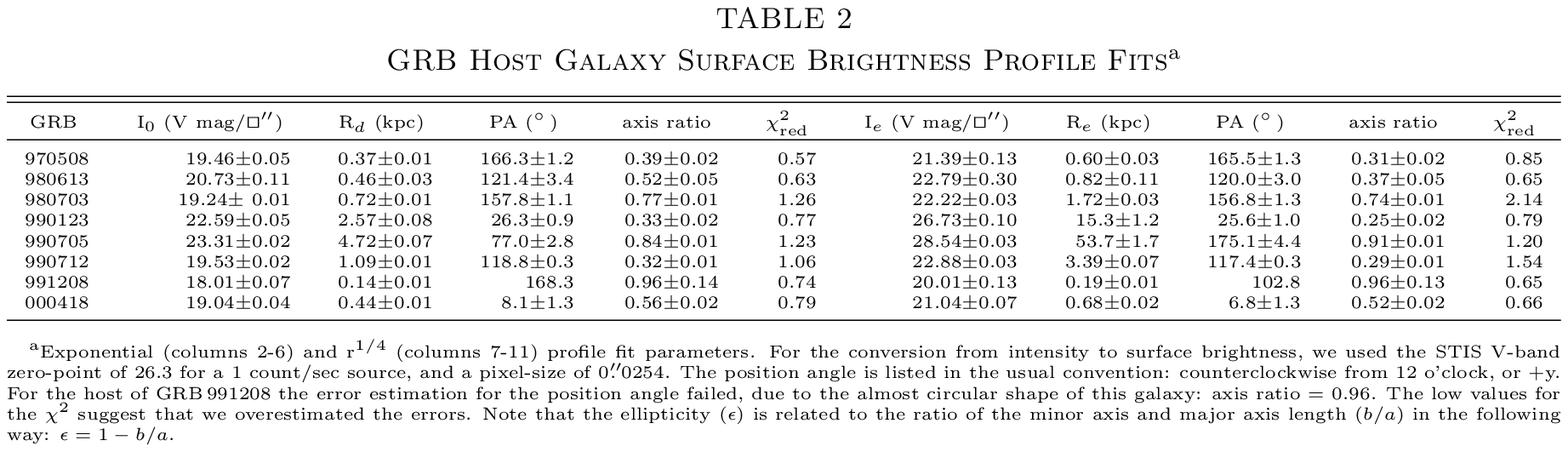}}
\end{center}
\vspace{-0.5cm}
\end{figure*}

\newpage

\normalsize


\begin{references}

\reference{} Abraham, R.G., Tanvir, N.R., Santiago, B.X., Ellis, R.S., Glazebrook, K., \& van den Bergh, S. 1996, MNRAS, 279, L47
\reference{} Barth, A.J., et al. 2003, ApJ, 584, 47L
\reference{} Berger, E., Cowie, L.L., Kulkarni, S.R., Frail, D.A., Aussel, H., \& Barger, A.J. 2003, ApJ, 588, 99
\reference{} Berger, E., Kulkarni, S.R., \& Frail, D.A. 2001, ApJ, 560, 652
\reference{} Bershady, M.A., Jangren, A., \& Conselice, C.J. 2000, AJ, 119, 2645
\reference{} Bertin, E., \& Arnouts, S. 1996, A\&AS, 117, 393
\reference{} Bloom, J.S., Djorgovski, S.G., \& Kulkarni, S. 2001, 554, 678
\reference{} Bloom, J.S., Kulkarni, S.R., \& Djorgovski, S.G. 2002, AJ, 123, 1111
\reference{} Bloom, S.J., Berger, E., Kulkarni, S., Djorgovski, S., \& Frail, D. 2003, ApJ, 125, 999
\reference{} Bloom, J.S., et al. 1999, Nature, 401, 453
\reference{} Budav{\' a}ri, T., Szalay, A.S., Connolly, A.J., Csabai, I., Dickinson, M. 2000, AJ, 120, 1588
\reference{} Bundy, K., Fukugita, M., Ellis, R.S., Kodama, T., \& Conselice, C.J. 2004, ApJ, 601, 123L 
\reference{} Bundy, K., Ellis, R., Conselice, C.J., 2005, ApJ, 625, 621
\reference{} Castro-Tirado, A., et al. 2001, A\&A, 370, 398
\reference{} Castro, S., Galama, T.J., Harrison, F., Holtzman, J., Bloom, J.S., Djorgovski, S., \& Kulkarni, S. 2003, ApJ, 586, 128
\reference{} Chapman, S.C., Windhorst, R., Odewahn, S., Yan, H., Conselice, C. 2003, ApJ, 599, 92
\reference{} Chary, R., Becklin, E.E., Armus, L. 2002, ApJ, 566, 229
\reference{} Christensen et al. 2004, astro-ph/0407066
\reference{} Christensen, L., Hjorth, J., \& Gorosabel, J. 2004, A\&A, 425, 913
\reference{} Conselice, C.J. 1997, PASP, 109, 1251
\reference{} Conselice, C.J., Bershady, M.A., \& Jangren, A. 2000a, ApJ, 529, 886
\reference{} Conselice, C.J., Bershady, M.A., \& Gallagher, J.S. 2000b, A\&A, 354, 21L
\reference{} Conselice, C.J., Gallagher, J.S., Calzetti, D., Homeier, N., \& Kinney, A. 2000c, AJ, 119, 79
\reference{} Conselice, C.J. 2003, ApJS, 147, 1
\reference{} Conselice, C.J., Bershady, M.A., Dickinson, M., \& Papovich, C. 2003a, AJ, 126, 1183
\reference{} Conselice, C.J., Chapman, S.C., Windhorst, R.A. 2003b, ApJ, 596, 5L
\reference{} Conselice, C.J., et al. 2004, ApJ, 600, 139L
\reference{} Conselice, C.J., Blackburne, J., \& Papovich, C. 2005, ApJ, 620, 564
\reference{} Daddi, E., et al. 2004, ApJ, 600, 127L
\reference{} Dickinson, M., et al. 2000, ApJ, 531, 624
\reference{} Djorgovski, S.G., Kulkarni, S.R., Bloom, J.S., Goodrich, R., Frail, D.A., Piro, L., \& Palazzi, E. 1998, ApJ, 508, 17L
\reference{} Djorgovski, S.G., Frail, D.A., Kulkarni, S.R., Bloom, J.S., Odewahn, S.C., \& Diercks, A. 2001, ApJ, 562, 654
\reference{} Djorgovski, S.G., Bloom, J.S., \& Kulkarni, S. 2003, 591, 13L
\reference{} Ferguson, H.C., et al. 2004, ApJ, 600, 107L
\reference{} Frail, D.A., et al. 2002, ApJ, 565, 829
\reference{} Franx, M., et al. 2003, ApJ, 587, 79L
\reference{} Fruchter, A.S., \& Hook, R.N. 2002, PASP, 114, 144
\reference{} Fruchter, A.S., et al. 2005, in prep
\reference{} Fruchter, A.S., et al. 2000, ApJ, 545, 664
\reference{} Fruchter, A.S., et al. 1999, ApJ, 516, 683
\reference{} Fruchter, A.S., Vreeswijk, P., Rhoads, J., \& Burud, I. 2001, GCN, 1200
\reference{} Fynbo, J.P.U., et al. 2003, A\&A, 406, 63L
\reference{} Galama, T.J., et al. 1998, Nature, 395, 670
\reference{} Galama, T.J., et al. 2000, ApJ, 536, 185
\reference{} Galama, T.J., et al. 2003, ApJ, 587, 135
\reference{} Gorosabel, J., et al. 2003, A\&A, 409, 123
\reference{} Heavens, A., Panter, B., Jimenez, R., \& Dunlop, J. 2004, Nature, 428, 625
\reference{} Hernandez-Toledo, H.M., Avila-Reese, V., Conselice, C.J., \& Puerari, I. 2005, AJ, 129, 682
\reference{} Hjorth, J., et al. 2003a, Nature, 423, 847
\reference{} Hjorth, J., et al. 2003b, ApJ, 597, 699
\reference{} Hjorth, J., et al. 2002, ApJ, 576, 113
\reference{} Holland, S., Bjornsson, G., Hjorth, J., \& Thomsen, B. 2000, A\&A, 364, 467
\reference{} Holland, S., et al. 2001, A\&A, 371, 52
\reference{} Hughes, D.H., et al. 1998, Nature, 394, 241
\reference{} Jha, S., et al. 2001, ApJ, 554, 155L
\reference{} Krist, J. 1995, in ASP Conf. Series 77, Astronomical Data Analysis Software and Systems IV, Vol. 4, 349
\reference{} Kuchinski, L.E., et al. 2000, ApJS, 131, 441
\reference{} Kulkarni, S.R., et al. 1998, Nature, 393, 35
\reference{} Kulkarni, S.R., et al. 1999, Nature, 398, 389
\reference{} Lehmer, B.D. et al. 2005, AJ, 129, 1
\reference{} Le Floc'h, E., et al. 2003, A\&A, 400, 499
\reference{} Le Floc'h, E., et al. 2002, ApJ, 581, 81L
\reference{} Lin, L. et al. 2004, ApJ, 617, 9L
\reference{} MacFadyen, A.I., \& Woosley, S.E. 1999, ApJ, 524, 262
\reference{} Masetti, N., et al. 2003, A\&A, 404, 465
\reference{} Metzger, M., et al. 1997, Nature, 387, 878
\reference{} Mirabel, N., et al. 2003, ApJ, 595, 935
\reference{} Mobasher, B., Jogee, S., Dahlen, T., de Mello, D., Lucas, R.A., Conselice, C.J., Grogin, N.A., Livio, M. 2004, ApJ, 600, 143L
\reference{} Moustakas, L., et al. 2004, ApJ, 2004, 600, 131L
\reference{} Nelder, J. \& Mead, R. 1965, Computer Journal, 7, 308
\reference{} Odewahn, S.C., et al. 1998, ApJ, 509, 5L
\reference{} Papovich, C., Dickinson, M., Giavalisco, M., Conselice, C.J., Ferguson, H.C. 2003, ApJ, 598, 827
\reference{} Petrosian, V. 1976, ApJ, 209, 1L
\reference{} Press, W.H., Teukolsky, S. A., Vetterling, W. T., \& Flannery,
B. P. 1992, Numerical Recipes in C (New York: Cambridge Univ. Press)
\reference{} Price, P., et al. 2002, ApJ, 571, 121L
\reference{} Reichart, D.E. 1999, ApJ, 521, L111
\reference{} Rhoads, J.E., \& Fruchter, A.S. 2001, ApJ, 546, 117
\reference{} Schade, D., Lilly, S.J., Crampton, D., Hammer, F., Le Fevre, O., \& Tresse, L. 1995, ApJ, 451, L1
\reference{} Soderberg, A.M., et al. 2004, ApJ, 606, 994
\reference{} Smith, I., et al. 2005, astro-ph/050357
\reference{} Somerville, R.S., et al. 2004, ApJ, 600, 135L
\reference{} Stanek, K.Z., et al. 2003, ApJ, 591, 17L
\reference{} Stanford, S.A., Dickinson, M., Ferguson, H.C., Lucas, R.A., Conselice, C.J., Budavari, T., \& Somerville, R. 2004, AJ, 127, 131
\reference{} Steidel, C.C., \& Hamilton, D. 1992, AJ, 104, 941
\reference{} Savaglio, S., \& Fall, F.M. 2004, ApJ, 614, 293
\reference{} Tanvir, N.R., et al. 2004, MNRAS, 352, 1073
\reference{} van den Bergh, S., Abraham, R.G., Ellis, R.S., Tanvir, N.R., Santiago, B.X., \& Glazebrook, K.G. 1996, AJ, 112, 359
\reference{} Vreeswijk, P.M., Fruchter, A., Ferguson, H., \& Kouveliotou, C. 2000, GCN, 751
\reference{} Vreeswijk, P.M., et al. 2001, ApJ, 546, 672
\reference{} Vreeswijk, P., Fruchter, A., Hjorth, J., Kouveliotou, C. 2003, GCN, 1785
\reference{} Vreeswijk, P.M., et al. 2004, A\&A, 419, 927
\reference{} Williams, R.E., et al. 1996, AJ, 112, 1335
\reference{} Windhorst, R.A., et al. 2002, ApJS, 143, 113
\reference{} Woosley, S.E. 1993, ApJ, 405, 273
\end{references}
\end{document}